\documentclass[amssymb,aps,twocolumn,showpacs]{revtex4}
\usepackage[]{graphicx}
\usepackage[]{verbatim}
\def\ket#1{| #1 \rangle}
\def\bra#1{\langle #1 |}

\begin{document}

\title{Experimental Quantum Communication without Shared Reference Frame}

\author{Teng-Yun Chen$^{1*}$, Jun Zhang$^{1*}$, J.-C. Boileau$^{2}$, Xian-Min
Jin$^{1}$,\\
Bin Yang$^{1}$, Qiang Zhang$^{1,3}$, Tao Yang$^{1,3}$ , R.
Laflamme$^{2}$, and Jian-Wei Pan$^{1,3}$}

\affiliation{
$^1$ Hefei National Laboratory for Physical Sciences at Microscale
and Department of Modern Physics, University of Science and Technology
of China, Hefei, Anhui 230026, China.\\
$^2$ Institute for Quantum Computing, University of Waterloo,
Waterloo, Ontario, N2L 3G1,Canada.\\
$^3$ Physikalisches Institut, Universit\"{a}t Heidelberg,
Philosophenweg 12, 69120 Heidelberg, Germany.}
\thanks{These authors contributed equally to this work.}
\date{\today}

\begin{abstract}

We present an experimental realization of a robust quantum
communication scheme [Phys. Rev. Lett. {\bf 93}, 0220501 (2004)]
using pairs of photon entangled in polarization and time. The
scheme overcomes errors due to collective rotation of the
polarization modes (for example, birefringence in optical fiber or
misalignment), is insensitive to phase's fluctuation of the
interferometer and does not require precise timing. No shared
reference frame is required except from the need to label the
different photons. We use this scheme to implement a robust
variation of the Bennett-Brassard 1984 quantum key distribution
protocol (BB84) over 1km of optical fiber. We conclude by
discussing and solving the unconditional security of our protocol.

\end{abstract}

\pacs{03.67.Dd, 03.65.Ud, 03.67.Pp}

\maketitle

Quantum cryptography~\cite{BB84}, whose security is based on the
fundamental principles of quantum mechanics, is a fast expanding
field of quantum information both theoretically and
experimentally~\cite{GRTZ02}. Recently, many quantum key
distribution (QKD) experiments have been realized through optical
fiber and free space using weak-coherent source or entangled
photon pairs.  The maximum distances of free space QKD using
weak-coherent source and entangled photons are 23.4km by
Kurtsiefer {\it et al.}~\cite{K02} and 13km by Peng {\it et
al.}~\cite{P05}, respectively. Their aim was to try to validate
the feasibility of quantum communication with satellites. Despite
some security flaws, fiber-based QKD over 100km has been
achieved~\cite{GYS04}.

Polarization and phase-time are most common coding methods to
implement QKD. Although polarization can be suitable for free
space QKD, it is generally not suitable for fiber-based QKD
because of the time and wavelength dependences of birefringence
which will depolarize the photons. Experimentally, active feedback
or self-compensation could be applied to solve these
problems~\cite{FJ1995}, but it is efficient only when the thermal
and mechanical fluctuations are rather slow. A popular alternative
to polarization coding is phase-time coding using unbalanced
interferometers~\cite{B92, HMP00}. However, phase-time coding can
be very sensitive to the phase's fluctuations between the two arms
of the interferometers and requires thermal stability. Some
ingenious tricks like two-way communication~\cite{MHHTZG1997} are
insensitive to phase's fluctuation, but have themselves disadvantages like being incompatible with perfect single photon sources and being sensitive to backscattering light.

To overcome the problems mentioned above, Walton {\it et al.}
proposed a scheme based on decoherence-free subspace (DFS) which
required encoding qubits using phase and time entanglement between
two photons~\cite{WASST2003}. Then Boileau {\it et
al.}~\cite{BLLM2004} proposed a variation of that protocol that
use a combination of time bins and polarization modes for coding.
These schemes are insensitive to phase's fluctuations of the
interferometer and robust against collective rotation induced by
birefringence or misalignment. In single photon QKD protocol, a
precisely synchronized clock is necessary to reduce the time
window to minimize the contribution of dark counts. However, it is
not the case for coding schemes using photon pairs, because the
photons simultaneously originating from the pair can provide
precise time references for each other. The fact that no
synchronized clock is necessary could be useful if the arrival
time of photon fluctuates.

The obvious disadvantages of two photon schemes are that they are
much more sensitive to photon loss and seem more inefficient than
the single photon schemes. However, it would be possible to reduce
the qubit losses to a level comparable to single photon schemes by
using post-selection, entanglement swapping and quantum memory
devices~\cite{BPTL2005}. As a step forward in that direction, we
implemented a variation of the BB84 protocol based on the robust
scheme of Boileau {\it et al.}~\cite{BLLM2004}, and realized an
efficient quantum communication without any shared reference
frame.

\begin{figure}[]
\centering
\includegraphics[width=86mm]{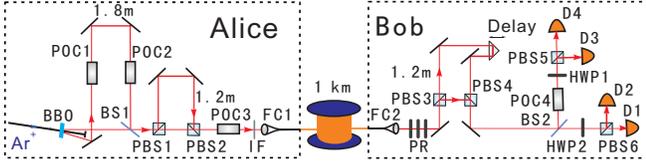}
\caption{\footnotesize{Experiment setup for robust QKD. A $400$ mW
Argon-ion laser beam ($Ar^{+}$) of $351$ nm passes through a $2$
mm beta-barium-borate (BBO) crystal and produces
polarization-entangled photon pairs of $702$ nm. POC1-POC4 are
four Pockel cells; BS1 and BS2 are beam-splitters; PBS1-PBS6 are
six polarized beam-splitters; FC1 and FC2 are two fiber couplers;
HWP1 and HWP2 are two half-wave plates (HWP); a HWP inserting
between two quarter-wave plates (QWP) constitutes a polarization
rotator (PR); IF is a interference filter; D1-D4 are four single
photon detectors.}} \label{Fig1}
\end{figure}

Our experimental scheme is illustrated in Fig.~\ref{Fig1}. On
Alice's side, polarization-entangled photon pairs are generated
via type-II spontaneous parametric down-conversion
(SPDC)~\cite{KMWZSS95}. The two photons of each pair are labelled
by passing two arms with 1.8m length difference. In the long arm,
two Pockel cells (POC1 and POC2), driven by high voltage pulse
generators gated with random number signal, are used to produce
the four states similar to that of BB84: $\ket{HV}+\ket{VH}$,
$\ket{HV}-\ket{VH}$, $\ket{HV}+i\ket{VH}$ and
$\ket{HV}-i\ket{VH}$, where $H$ and $V$ stand for horizontal and
vertical polarization mode, respectively.

The two entangled photons can be combined into the same path in
the beam-splitter BS1 with a probability of $1/4$. Afterward, the
vertically polarized photons are firstly tagged with a delay $T$
in an unbalanced interferometer composed of two polarizing
beam-splitters (PBS1 and PBS2) with a 1.2m difference between the
two pathes' length. The POC3 is added to perform a random
collective rotation of polarization before the photons are
collected into the fiber coupler (FC1). Then, the photons are sent
to Bob directly or through a 
1 km single mode optical fiber. A
polarization rotator (PR) is used to simulate the collective
rotation noise. On Bob's side, the received horizontally polarized
photons are tagged with the same delay $T$. To make the two timing
tags exactly the same, a right-angled prism (Delay in
Fig.~\ref{Fig1}) is used to adjust the path length precisely.

Using notation introduced in Ref.~\cite{BLLM2004} and supposing
that the initial state was of the form
$\alpha\ket{HV}+\beta\ket{VH}$, the resulting state can be written
as:
\begin{eqnarray}
&& ((\delta_1+1)/2)(\alpha\ket{H'_TV'_T}+\beta \ket{V'_TH'_T})\nonumber\\
&& +((\delta_1-1)/2)(\alpha\ket{V'H'_{TT}}+\beta\ket{H'_{TT}V'})\nonumber\\
&&+((\delta_2 +\delta_3)/2) (\alpha \ket{H'_TH'_{TT}}+\beta \ket{H'_{TT}H'_T})\nonumber\\
&&+((\delta_2 -\delta_3)/2) (\alpha \ket{V'V'_T}+\beta
\ket{V'_TV'}),\nonumber
\end{eqnarray}
\label{eq1}where $\ket{H'}$ and $\ket{V'}$ represent Bob's
polarization basis frame which can be different from Alice's one.
The subscripts T and TT mean that the photon has been tagged once
and twice, respectively. The $\delta_j$'s are parameters that
depend directly on the collective rotation of the polarization
mode. They satisfy the following relation:
$||\delta_1||^2+||\delta_2||^2+||\delta_3||^2=1$ . Giving the
arrival time of the photons, the final state is projected to the
original state with a probability
$p_s=||\frac{\delta_1+1}{2}||^2$. The $p_s$ could be anything
between 0 and 1. To make $p_s$ independent of the environment or
any misalignment of the reference frame, Alice or Bob could apply
a random unitary transformation $B^{\otimes2}$ between the two
tagging operations. If $B$ is chosen from the uniform distribution
over $U(2)$, then $p_s$ is in average equal to $\frac{1}{3}$
whatever is the collective rotation. Because it's difficult to
realize random transformation $B$ over the whole $U(2)$ space
experimentally, we simplify the experimental set-up by using only
one POC (POC3 in Fig.1). 
Making it do nothing half of the time, and a bit-flip operation otherwise,
$p_s$ could also
average to a non-zero value, $\frac{1}{4} \leqslant p_s \leqslant
\frac{1}{2}$.


The received photons are split at BS2. The two half-wave plates HWP1 and HWP2 are set such that they performs as Hadamard gates on the polarization. By switching POC4 such that it do nothing or act as a QWP at $90{^{\circ}}$, we can select a random measurement basis (either  $\{\ket{H'_TV'_T}+\ket{V'_TH'_T},\ket{H'_TV'_T}-\ket{V'_TH'_T}\}$ or the $\{\ket{H'_TV'_T}+i\ket{V'_TH'_T},\ket{H'_TV'_T}-i\ket{V'_TH'_T}\}$).  By post-selecting the cases where each of two photons exit from different outputs of BS2 and their arrival time difference is $6$ns (which is related to the $1.8$m time label), the states are differentiated according to their polarization (the same or different) ~\cite{BLLM2004}. The detection events within the $3$ns coincident time window are recorded to generate quantum key bits.

\begin{figure}[]
\begin{minipage}[c]{.24\textwidth}
\flushleft
\includegraphics[scale=0.55]{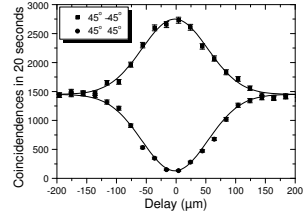}
\end{minipage}
\begin{minipage}[c]{.23\textwidth}
\caption{\footnotesize{Interference pattern observed for the state
$\ket{H'_TV'_T}-\ket{V'_TH'_T}$ after applying a Hadamard gate on
each photon and by measuring the coincident counts. The zero delay
position corresponds to the maximum interference visibility.}}
\label{Fig2}
\end{minipage}
\end{figure}

For the measurement to succeed, it is crucial to observe two
photons interference after the timing tags.  It requires to match
accurately the difference of the path's lengths of the two
interferometers by adjusting the prism on Bob's side (see
Fig.~\ref{Fig1}). The curve in Fig.~\ref{Fig2} shows an
interference fringe with a visibility of above $95\%$. The fact
that interference is observed over a large length interval (of
about one hundred micrometers) clearly implies that the
interference is robust against the phase instability of the
interferometers as claimed in Ref.~\cite{WASST2003, BLLM2004}.

In order to demonstrate the robustness of the protocol in
principle, we first use a $4$m optical fiber to implement the QKD
protocol. Approximately $12,000$Hz polarization-entangled pairs
are detected behind a interference filter (IF) of $1.6$nm FWHM.
The entangled photon pairs are transferred to one of the four
states randomly and sent to Bob. Due to the photon losses in the
BSs and the fiber connecters, only a maximum of $140$Hz coincidences can be
registered on Bob's side after calibrating the PR. 
We then rotate the angle of the first QWP of the PR to simulate
the degree of collective rotation noise. In the experiment, five
settings are selected for particular angles of the QWP. The first
setting corresponds to the case where there is no collective rotation and coincidence is maximal. The last setting corresponds to a collective bit-flip. 
The other settings are chosen via rotating the angle of QWP with equal
intervals between the best and the worst settings. We investigated
the change of error rates and coincidences under these conditions
with or without random rotation implemented by POC3.

\begin{figure}[ht!]
\flushleft
\includegraphics[scale=0.7]{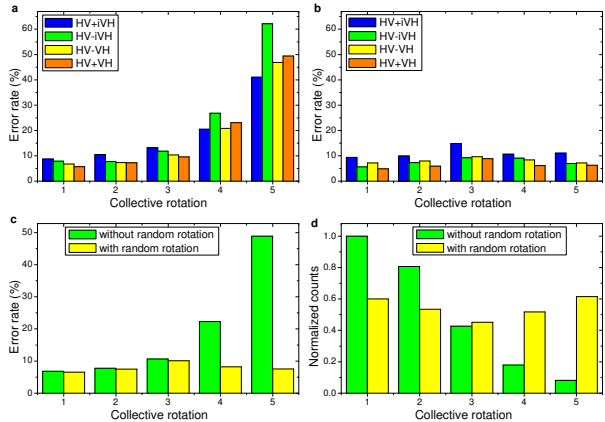}
\caption{\footnotesize{
Quantum bit error rates (QBER) measured under different collective rotations
and with a $4$m single mode fiber. The first collective rotation correspond to the identity and the fifth, to a bit-flip. The angles in between are chosen has described in the text. The QBER of each states was measured in 20 minutes without (a) or with random rotations (b). The average QBER over all states with and without the random rotations are compared in (c). In (d), we give the normalized coincidence counts in function of the angle of the collective rotation with or without random rotations.
}} \label{Fig3}
\end{figure}


As shown in Fig.3, the coincidence without random rotation is very
dependent of the collective noise. When
the angle of rotation increases, the coincidence decreases
to a minimum while the error rate increases close to $50\%$.
As predicted, using random rotation makes the coincidence and the error rates much less
dependent on the collective noise. We later show that the all the error rates obtained with the random rotations are suitable for secured QKD.

We also constructed a practical QKD system over
$1$km single mode fiber, whose attenuation for each photon at 702
nm is $4.8$ db/km. Due to the photon loss in the fiber and an
additional fiber connecter, the maximal coincidence detected in
Bob's side dramatically drops to $1.4$Hz. 
We measured the error rates under the same
collective noise settings as used in the experiment with short
fiber. The results of the experiment are shown in Fig.4. Due to the photon loss, the QBER observed in the experiment is a bit higher than the cases with short fiber, but is still well below the lower bound for secure key distribution.

\begin{figure}[ht!]
\includegraphics[scale=0.7]{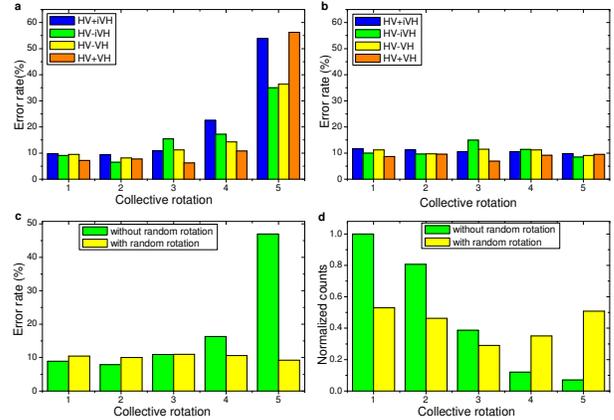}
\caption{\footnotesize{QBER measured under different collective rotations (as in Fig.~\ref{Fig3})
and with a $1$km single mode fiber. The QBER of each states was measured in 3 hours without (a) or with random rotations (b). The average QBER over all states with and without the random rotations are compared in (c). In (d), we give the normalized coincidence counts in function of the angle of the collective rotation with or without random rotations.
}} \label{Fig4}
\end{figure}

In fact, the error rates are mostly come from the imperfection of
state preparation and accidental coincidence. The 2000 Hz single
count rate of each detector and the 3ns coincidence window lead to
an accidental coincidence of $0.024$Hz. When the collective noise
changes, the accidental coincidence will induce an error rate from
$1.7\%$ to $50\%$ when no random rotation is applied. However,
with the help of random rotation it will only cause an error rate
varied from $3.4\%$ to $6.8\%$, which is much more independent of
the collective noise. The imperfect state source will also cause
an error rate of $4\%$ . The average error rate is observed to be
$10.2\%\pm0.3\%$ (see Fig.4-c). The following security proof will
show that our experimental method can successfully distribute
quantum key over a practical collective noise channel without
shared reference frame.

It should be noted that our experimental measurements are not exactly the
same as the ones required by the standard BB84 protocol. However,
if Bob was able to do a projection into the space $S$ spans by
$\ket{H'V'_T}$ and $\ket{V'_TH'}$, then both measurements would be
identical. Instead of doing this projection directly, suppose
that, just before his tag operation, Bob apply on the polarization
modes of the photon pairs a random phase shift $M_\phi= \left(
\begin{array}{cc}
 1 &  0   \\
 0 &  e^{i\phi}
\end{array}
\right) ^{\otimes2}$ with $\phi$ chosen between 0,
$\frac{\pi}{2}$, $\pi$ or $\frac{3\pi}{2}$.

Because of the post-selection, the only states that we need to
consider are the ones of the form
$a_1\ket{H'V'_T}+a_2\ket{V'_TH'}+a_3\ket{H'H'}+a_4\ket{V'_TV'_T}$
for some complex numbers $a_i$. Consider the density matrix $\rho$
of an arbitrary mixture of these states. Simple calculations show
that the elements of $\rho'=\sum_{\phi=0,\frac{\pi}{2},\pi,
\frac{3\pi}{2}} M_\phi \rho M_\phi$ that corresponds to
$\ket{V'_TV'_T}\bra{H'V'_T}$,
$\ket{H'H'}\bra{H'V'_T}$,$\ket{V'_TV'_T}\bra{V'_TH'}$,
$\ket{H'H'}\bra{V'_TH'}$ and their transposes vanish. In other
word, sometime the state is projected into or outside $S$. In the
first case, the measurement reduce to one of the two von Neumann
measurement used in the standard BB84. In the other case, the
state may be projected outside $S$ and the measurement fails. If
Alice and Bob were able to know exactly which pairs were projected
inside or outside $S$ then they could reject those pairs that
correspond to the wrong projection and could perform the standard
BB84 protocol with the remaining pairs. However, they don't have
this information. Instead, they can measure with a good accuracy
the ratio of the states that are projected inside $S$, which we
refer to as $p^S$. This can be achieved by selecting a random
sample of photon's pairs on which the measurement corresponding to
$\{ \ket{H'_TH'_T},\ket{H'_TV'_T}, \ket{V'_TH'_T}, \ket{V'_TV'_T}
\}$ is applied immediately after the tagging operation, and by
counting the results corresponding to either $ \ket{H'_TV'_T}$ or
$\ket{V'_TH'_T}$. If that sample is large enough, the measured
ratio of states projected inside and outside $S$ will be very
close to $p^S$. Remark that the number of pairs required for the
measurement of $p^S$ is asymptotically negligible.

To obtain a secure key, it is necessary and sufficient to bound Eve
information about the key after bit error correction since privacy
amplification~\cite{BBR88, RK05} can be used to reduce
asymptotically that information to zero with a key's lost
proportional to Eve's information. For the qubits projected outside
$S$, Alice and Bob assume the worst case scenario and suppose that
Eve has full information about the results corresponding to these
states. For the qubits projected inside $S$, Shor and Preskill's
proof~\cite{SP00} bound Eve's information after bit error correction
by $H(e_x^{\rm S})$. Consequently, the secret key generation rate is
at least $p^S-H(e_x)-p^SH(e_x^S)$ of the conclusive results. Note
that, H is the Shannon entropy, $e_x$ and $e_x^S$ are the bit error
rate over all conclusive results and over the conclusive results
that were projected inside S, respectively. A conclusive result is
defined as any measurement that gives a bit to the key before error
correction and privacy amplification.

Since both $e_x$ and $p^S$ can be measured directly by using a
sample of test bits, to estimate the secret key generation rate,
Alice and Bob only need an upper bound for $e_x^S$. $e_x^S$ can be
estimated from the identity $e_x=p^Se_x^S+(1-p^S)e_x^{\bar{S}}$,
where $e_x^{\bar{S}}$ is the error rate of the conclusive results
corresponding to the states projected outside S. As a consequence of
Bob¡¯s random choice of $\Phi$ for $M_{\Phi}$ and the fact that the
coefficients of the density matrix $\rho'$ corresponding to
$\ket{V'_TV'_T}\bra{H'H'}$and $\ket{H'H'}\bra{V'_TV'_T}$ are zeros,
$e_x^{\bar{S}}$ asymptotically. In the experiment, we measured $p^S$
using the method explained above. In the case with 4 m fiber, $p^S$
is measured to be 97\% and in the case with 1 km fiber $p^S$ is
91\%. With the help of random rotations, the $e_x$¡¯s observed in
both cases (see Figs.\ref{Fig3} and \ref{Fig4}) are sufficient to
guarantee secure key distribution.

Any coherent attack from an eavesdropper was considered in our
security analysis. However, we assumed perfect state preparation and
measurements, and that Eve's has no access whatsoever to Alice and
Bob's lab. For a more realistic security analysis, considerations as
the ones treated in Ref.~\cite{GLLP02, GFKZR05} would be necessary.

In summary, we have realized one of the first efficient quantum
communication protocols without shared spatial and reference frame
including no time reference, except to label the qubits. It could be
useful for free-space transmission in the case where the receiver
and the sender are moving relative to each other. It could also be
useful to avoid birefringence effect in optical fiber and would be a
possible solution to the phase instability of interferometers. Our
experiment is a first step toward more efficient robust quantum
communication since it is only an example of a series of more
complex quantum communication schemes exploiting the
decoherence-free subsystem of the collective noise and time
tags\cite{BK05}.We also showed the unconditional security of a
robust quantum key distribution protocol based of BB84. We conclude
with the remarks that technological advances of entangled photon
sources and quantum memories would greatly enhance our results.

This work was supported by the CAS, the NNSFC, the NSERC and the
PCSIRT. This work was also supported by the Alexander von Humboldt
Foundation, the Marie Curie Excellence Grant of the EU and the
government of Ontario.


\begin{thebibliography}{}

\bibitem{BB84} C. H. Bennett and G. Brassard, in {\it Proceeding of the IEEE
International Conference on Computers, Systems, and Signal Processing, Bangalore},
India (IEEE, New York, 1984), pp.175-179 (1984).

\bibitem{GRTZ02} N. Gisin, G. Ribordy, W. Tittel and H. Zbinden, {\it Rev. Mod. Phys.} {\bf 74}, 145 (2002).

\bibitem{K02}  C. Kurtsiefer {\it et. al.}, Nature (London) {\bf 419}, 450 (2002).

\bibitem{P05} C.-Z. Peng {\it et. al.}, Phys. Rev. Lett. {\bf 94}, 150501 (2005).

\bibitem{GYS04}  C. Gobby, Z. L. Yuan and A. J. Shields, {\it Appl. Phys. Lett.} {\bf 84}, 3762 (2004); T. Kimura {\it et. al.}, {\it quant-ph/0403104}.

\bibitem{FJ1995} J.D. Franson and B.C. Jacobs, {\it Electron. Lett.} {\bf 31}, 232 (1995).

\bibitem{B92}  C. H. Bennett, {\it Phys. Rev. Lett.} {\bf 68}, 3121 (1992).

\bibitem{HMP00}  R. Hughes, G. Morgan and C. Peterson, {\it J. Mod. Opt.} {\bf 47}, 533
(2000); G. Ribordy {\it et. al.}, {\it ibid.} {\bf 47}, 517
(2000); W. Tittel, J. Brendel, H. Zbinden and N. Gisin, {\it Phys.
Rev. Lett.} {\bf 84}, 4737 (2000).

\bibitem{MHHTZG1997}  A. Muller {\it et. al.}, {\it Appl. Phys. Lett.} {\bf 70} 793(1997);
D. Stucki {\it et. al.}, {\it New J. Phys.} {\bf 4}, 41 (2002).

\bibitem{WASST2003} Z.D. Walton {\it et. al.}, {\it Phys. Rev. Lett.} {\bf 91}, 087901 (2003).

\bibitem{BLLM2004} J.-C. Boileau, R. Laflamme, M. Laforest and C. R. Myers, {\it Phys. Rev. Lett.} {\bf 93}, 220501 (2004).

\bibitem{BPTL2005} J.-C. Boileau, D. Poulin, K. Tamaki and R. Laflamme, {\it to be published).}.


\bibitem{KMWZSS95} P.G. Kwiat {\it et. al.}, {\it Phys. Rev. Lett.} {\bf 75}, 4337 (1995).



\bibitem{BBR88} C. H. Bennett, G. Brassard and J.-M. Robert, {\it SIAM Journal on Computing}, 17(2): 210, (1988).

\bibitem{RK05} R. Renner and R. K\"{o}nig, In {\it Theory of Cryptography-TCC 2005, LNCS(Springer, Berlin, 2005)}, p.407.


\bibitem{SP00} P. W. Shor and J. Preskill, {\it Phys. Rev. Lett.} {\bf 85}, 441 (2000).

\bibitem{GLLP02} D. Gottesman,  H.-K. Lo, N. L\"{u}tkenhaus and J. Preskill, {\it Quant. Info. and Comp.} Vol.{\bf 4}, No. 5, 325 (2004).

\bibitem{GFKZR05} N. Gisin {\it et. al.}, {\it Phys. Rev. A} {\bf 73}, 022320 (2006).

\bibitem{BK05} J. L. Ball and K. Banaszek, {\it Open Syst. Inf. Dyn.} {\bf 12}, 121 (2005).




\end{thebibliography}
\end{document}